\documentstyle[12pt,aaspp,epsf,epsfig]{article}

\def\lsim{\lower.5ex\hbox{$\; \buildrel < \over \sim \;$}}
\def\gsim{\lower.5ex\hbox{$\; \buildrel > \over \sim \;$}}

\def\li#1{\hbox{$^{#1}{\rm Li}$}}
\def\b1#1{\hbox{$^{1#1}{\rm B}$}}
\def\be#1{\hbox{$^{#1}{\rm Be}$}}
\def\msun{${\,M_\odot}$}
\def\etal{{\it et al.}~}
\def\beginapjbib{\begingroup \section*{\large \bf References}
   \parskip=.5ex plus 1.0pt
   \def\bibitem{\par \noindent \hangindent\parindent
      \hangafter=1}}
\def\endapjbib{\par \endgroup}

\def\beq{\begin{equation}}
\def\eeq{\end{equation}}

\begin{document}
\rightline{UMN-TH-1706/98}
\rightline{astro-ph/9806084}
\rightline{June 1998}
\title{Testing the Primary Origin of\\
Be and B in the Early Galaxy}

\author{Elisabeth Vangioni-Flam}
\affil{Institut d'Astrophysique de Paris\\ 98bis
Boulevard Arago\\ 75014 Paris, France}

\author{Reuven Ramaty}
\affil{Laboratory for High Energy Astrophysics\\ Goddard Space 
Flight Center\\ Greenbelt, MD 20771}

\author{Keith A. Olive}
\affil{School of Physics and Astronomy\\ University of Minnesota\\
 Minneapolis, MN 55455, USA}

\and

\author{Michel Cass\'e}
\affil{Service d'Astrophysique\\ DSM, DAPNIA, CEA,  France \\
and \\ Institut d'Astrophysique de Paris\\ 98bis
Boulevard Arago\\ 75014 Paris, France}

\begin{abstract}

Beryllium and boron measurements in metal poor stars have had a major 
impact on our understanding of the origin of the light elements in the
universe. Two types of
models have been proposed to explain the linear rise of 
the Be and B
abundances as a function of iron observed in metal poor halo stars. In
both cases, this linearity indicates that freshly synthesized C
and O are
accelerated by Type II supernovae and subsequently fragmented into Be
and B.  One mechanism advocates shock acceleration in the gaseous
phase of superbubbles excavated by collective SNII explosions. 
Because of their short lifetimes, only the most massive stars (with
an initial mass greater than 60\msun) do not drift out of
superbubbles, and  participate in BeB production. The second
mechanism is based on
 the acceleration of the debris of grains formed in the ejecta of
all SNIIs (originating  from stars with initial mass greater than
8\msun). Here again, fresh C and O are sped up to cosmic ray energies
by shocks.

We propose a possible test to discriminate between the two 
scenarios. If supernovae of all masses are involved in BeB 
production, the Be/Fe ratio is constant, since both elements are 
produced in the same events. Alternatively, when only the most 
massive stars are involved in Be production, Be/Fe is enhanced at 
very early times because of the shorter lifetimes of these stars. 
This predicted difference in the
behavior of Be/Fe could be tested by high quality observations at [Fe/H]
$\lsim -3$.

We also note that the solution invoking only the most massive
supernovae  mimics a flat evolution of both Be/H and B/H as a
function of Fe/H at low metallicity, and could thus resemble a
``plateau" for these elements despite a lack of a primordial Big Bang
nucleosynthesis origin. Consequently, there may be no need to invoke
inhomogeneous Big Bang  models to explain the initial production of
BeB should a plateau be discovered. 
\end{abstract}


\newpage

\section{Introduction}

The origin and evolution of light elements (\li6, \li7, \be9, \b10, 
and \b11) is an important chapter in the development of nuclear 
astrophysics. In the 70's and 80's the problem of Li, Be and B 
(hereafter LiBeB) nucleosynthesis has been considered essentially 
solved by Galactic Cosmic Ray (GCR) spallation (Meneguzzi, Audouze 
\& Reeves 1971, Reeves 1994). The constituent nuclei of 
the
GCRs, protons 
and $\alpha$ particles, as well as C, N and O (hereafter CNO), form 
LiBeB via spallation on stationary nuclei in the interstellar medium 
(ISM). \li7 which has additional sources of production 
is an exception, as is \b11  since the \b11/\b10 isotopic ratio is 
not correctly predicted by GCR nucleosynthesis. An artificial 
solution for the B isotopic ratio had been proposed, based on a 
non-observable low energy spike in the GCR energy spectrum, the 
so-called ``carrot" (Meneguzzi \& Reeves 1975).

New observations in the late 80's prompted a reassessment of the 
question as to the origin of LiBeB.  Be abundance measurements in 
halo stars were achieved down to [Fe/H] = -1.5 (Rebolo \etal 1988, 
Ryan \etal 1990). As is generally the custom, square brackets will 
denote logarithmic abundance ratios by number relative to the solar
value. A good fit of the Be  evolution was obtained within the limited
range of these  observations (Vangioni-Flam \etal 1990) by considering
the  progressive CNO enrichment of the ISM due to stellar production and 
injection throughout the lifetime of the Galaxy, and supposing that 
the GCR flux is proportional to the SN rate (SN shocks serving only 
to accelerate particles out of matter of the same metallicity as 
that of the interstellar medium). At that time, these evolutionary 
effects on both GCR nucleosynthesis and the ISM, were sufficient to 
explain the behavior of Be vs. Fe.  Subsequently, however, data were 
obtained at even lower metallicities for beryllium (Gilmore \etal 
1992, Ryan \etal 1994, Boesgaard \& King 1993) and a few boron 
abundance measurements were made over a wide metallicity  range 
(Duncan \etal 1992, Edvardsson \etal 1994). These observations 
indicated a quasi linear relationship between both Be and B vs. Fe, 
instead of the quadratic relationship expected if the GCR were 
accelerated out of the ISM. This increased the general perplexity of 
potential solutions (Pagel 1991) and gave rise to a new wave of 
research (Duncan \etal 1992, Walker \etal 1993, Feltzing \& 
Gustaffson 1994, Vangioni-Flam \etal 1994, Cass\'e \etal 1995, 
Fields \etal 1995, Bykov 1995,
Ramaty \etal  1996, Vangioni-Flam \& Cass\'e 
1996). The primary origin of beryllium and boron (i.e. 
the fact that the production rate is independent of the ISM 
metallicity) indicates that these elements result from the 
spallation of fresh products of nucleosynthesis  (primarily from C 
and O), rather than nuclei accumulated in the ISM. Thus, we are 
presented with the challenge to find an appropriate mechanism 
different from the traditional GCR picture which has become 
problematic for two reasons. First, as we noted above, if the cosmic 
rays are accelerated out of the ISM and interact in the ISM, the 
rising CNO/H abundance in the ISM leads to cumulative Be and B 
abundances which depend quadratically on the ISM metallicity, and 
thus is in disagreement with the observations. In addition, Be 
production in the early Galaxy by GCR accelerated out of the ISM 
requires the supply of extraordinarily large amounts of energy to 
the cosmic rays (Ramaty \etal 1997; Ramaty, Kozlovsky \& 
Lingenfelter 1998). 

The carrot of Meneguzzi \& Reeves (1975) (introduced to explain \b11),
having the  same GCR  composition, is also problematic  because the B 
production  by low energy cosmic rays should give rise to a 
quadratic relationship instead of a linear one exactly as in the 
high energy case.  Moreover on theoretical grounds, a low energy 
spike throughout the galactic history would lead to an 
overproduction of Be and B (Lemoine \etal 1998). Finally, Li would 
be overproduced in the early galaxy by the $\alpha + \alpha$ 
reactions, spoiling the observed (primordial) Li plateau. 

Accelerated particle reactions are not the only sources of boron 
since carbon spallation by neutrinos in core collapse supernovae 
(Types II and Ib, hereafter SNII) can also contribute significantly 
to \b11 production (Woosley \etal 1990; Olive \etal 1994, Woosley 
\&Weaver 1995, Vangioni-Flam \etal 1996; Ramaty \etal 1997). This 
mechanism is particularly interesting because it yields mainly 
$^{11}$B, making $\nu$-induced spallation important for the 
explanation of the meteoritic B isotopic ratio, $^{11}$B/$^{10}$B = 
4.05$\pm$0.2 (Chaussidon \& Robert 1995). Vangioni-Flam \etal (1996) 
found that the neutrino contribution to the total B production 
should amount to at most $\sim$30\%. If the cosmic ray spectrum 
extends to high energies, \b11 production by neutrino spallation is 
always required since such cosmic rays are not capable of 
reproducing the meteoritic ratio, but again a $\sim$30\% 
contribution to the total B production from neutrinos appears 
sufficient (Ramaty \etal 1997).

As the spallation of C, N and O is the only significant source of 
Be, the linear dependence of [Be/Fe] with respect to [Fe/H] (at 
least up to [Fe/H] = -1), or equivalently the approximate constancy 
of [Be/Fe], implies that a mechanism whereby C and O are accelerated
above the  spallation thresholds and impinge on the ambient H and He is
operative. SNII's are the most plausible sources of accelerated C and O
in  the early Galaxy. Accelerated N, however, makes only a very 
minor contribution since it is highly underproduced in SNII's. Among  the
different scenarios proposed to explain the linear evolution of  Be and
B, we consider the following two, which shall subsequently be  referred
to as models (a) and (b). 

In model (a) Be and B are produced both by low energy nuclei
(hereafter LEN), highly enriched in C and O 
relative to H and He, and standard GCR accelerated out of the ISM 
(Cass\'e, Lehoucq \& Vangioni-Flam 1995). The latter is only dominant at
late times in the evolution of the Galaxy.  This model was motivated
by the observations of a linear dependence of Be and B on Fe
which implies a primary source for their production and
the observations of C and O deexcitation gamma ray
line emission  from Orion (Bloemen et al. 1994; 1997).  It was 
suggested (Bykov 1995, Parizot \etal 1997) that the required 
population of C and O enriched LEN  could result from the 
acceleration, by an ensemble of weak shocks in superbubbles, wherein 
the seed particles for acceleration originate from the winds of 
massive stars and the ejecta of supernovae from massive star 
progenitors. Only the most massive stars (M$>$60\msun), that is 
those which explode within superbubbles due to their very short 
lifetime, should be involved in this scenario.  In the early Galaxy,
these extended acceleration sites would be sustained essentially by SNII
exploding in OB associations (Samland 1998). Later on, in the disk phase,
WR stars would also participate, since the stellar winds  intensify at
increasing metallicities (Meynet \etal 1994). The scenario further 
assumes that the metallicity of the LEN component is independent of 
the average Galactic metallicity, thereby dominating the Be 
production in the halo phase ([Fe/H] $\lsim$ -1 with the GCRs taking 
over in the disk phase (Vangioni-Flam \etal 1996, 1997). 

In model (b) Be and B are produced by standard GCRs accelerated at 
all epochs of Galactic evolution from the ejecta of supernovae 
(Ramaty \etal 1997; 1998). This model, motivated by the observed, 
essentially constant [Be/Fe] in the early Galaxy, envisions the 
acceleration of the erosion products of high velocity refractory 
grains formed in a supernova ejecta (Lingenfelter, Ramaty \& 
Kozlovsky 1998). These 
authors have shown that sufficient O is incorporated in refractory 
Al$_{2}$O$_{3}$, MgSiO$_{3}$, Fe$_{3}$O$_{4}$ and CaO to account for 
the GCR source O abundance. They have further argued that the GCR 
source C abundance could also be understood if the fraction of 
ejecta C incorporated in refractory grains (mainly graphite) is the 
same as that of the other main refractories, and they have shown 
that the standard arguments against the acceleration of the 
refractory metals out of supernova ejecta are model dependent and 
answerable in principle. It is thus possible that at all epochs of 
Galactic evolution the standard GCR would contain sufficient C and O 
to explain the linear Be evolution. In this scenario, individual 
SNII with progenitors of the same mass range as that responsible for 
Fe production (M$>$8\msun) participate in the production of Be and 
B. 

If the Be in the early Galaxy is indeed produced by particles whose 
acceleration is related to short-lived very massive stars, then the 
difference in the lifetimes of the progenitors of Be and Fe and the 
relative  number of stars implied in each case, could also affect 
the evolution of Be/Fe. In the present paper we shall critically 
examine the evolution of B and Be in the early Galaxy, taking into 
account: i) the relative Be yields associated to each mass domain 
considered and  ii) potential time dependent effects due to the 
mass dependence of the lifetimes of the stellar progenitors of the 
core collapse supernovae responsible for the production of B and Be. 
In the  following, we reproduce the observed Be evolution through a
Galactic evolutionary model and explore the correlated behavior of 
B considering three plausible B/Be production ratios, in agreement 
with the results of the  nuclear spallation models of various  
compositions and energy spectra (Vangioni-Flam \etal 1996, Ramaty 
\etal 1997).  The wide range of B/Be ratios explored leaves room for 
neutrino spallation. 

In what follows,  we will examine whether or not it is possible,  
using the existing data on B and Be, to distinguish between models 
a) and b). In section 2, we will describe the  current status of the 
B and Be data. In section 3, we will describe and develop the 
proposed test to distinguish between the models and present the 
results of our calculations. Our conclusions are found in section 4.

\section{Data}

There is a three-fold advantage in studying 
Be. First,  abundant data exist over a large range of 
 metallicities. It is a pure spallation product not contaminated by
 neutrino-spallation as is \b11 and \li7 or the stellar production as
 \li7. And finally, its measured abundance is not significantly altered
by NLTE effects which in the case
of B are difficult to estimate. For these reasons, we will focus
primarily on Be.

The last decade has seen considerable progress since the 
early observations of Rebolo \etal (1988) and 
Ryan \etal (1990) of a total of
ten low metallicity halo dwarf stars, all yielding upper limits
with three potential determinations of a Be abundance.
Since then, there have been at least 50 new observations of 25 additional
halo dwarfs (Gilmore \etal 1992, Boesgaard \& King 1993, 
Ryan \etal 1994, Primas 1995, Rebolo \etal 1993,
Garcia-Lopez, Severino, \& Gomez 1995, Thorburn \& Hobbs 1996,
Molaro \etal 1997).  We have compiled the Be data from the literature
and show the Be abundances as a function of [Fe/H] in Figure 1.
The data have been combined systematically so that each point 
corresponds to a single star.  Where multiple observations of
a star are found, the Be abundances are first adjusted by taking a 
common set of stellar parameters (surface temperature, surface gravity, 
and metallicity) followed by a weighted average of the different 
observations. When possible, we have assumed temperatures
as given by Fuhrmann, Axer, \& Gehren (1993). 
For example, we have found seven distinct measurements
of HD 140283.  The Be abundances range from $\log$(Be/H) = -13.25 to -12.85
with assumed surface gravities running from 3.2 to 3.56, temperatures
from 5540 to 5814, and metallicities from -2.2 to -2.77.
Here, we have taken  g = 3.4,  T = 5814, and [Fe/H] = -2.6, which reduces
the range for $\log$(Be/H) to -13.11 to -12.87 and yields an average
$\log$(Be/H) = -12.97 $\pm$ 0.07 for this star.

The large number of Be observations in low metallicity halo stars have
shown that the Be/H abundance ratio increases approximately linearly with
[Fe/H] up to at least one tenth of the solar metallicity.
Because of the multiple observations of many of the halo dwarfs,
the errors in the determined Be abundances are relatively small.
In contrast, the Fe abundances are particularly uncertain, and in one
case, the assumed values of [Fe/H] differ by as much as $\sim$ 0.6
dex. As a conservative estimate for the error in  [Fe/H],
we have taken 0.2 dex. A linear regression on the
data for $\log$(Be/H) vs. [Fe/H] (for [Fe/H] $<$ -1) then yields 
\beq
\log ({\rm Be/H}) = (-10.03 \pm 0.18) + (1.18 \pm 0.10) {\rm [Fe/H]}
\label{be}
\eeq
Clearly, this regression indicates a predominantly primary origin for
beryllium (secondary Be would give a slope of 2 rather than 1.18 as in 
eq. (\ref{be}).
As yet, the data show no signs of revealing a plateau which could be 
interpreted as a primordial value for Be as in the case of Li (though see
below for a complication on this interpretation).
This is of course not a surprise since in standard big bang nucleosynthesis
calculations the primordial value of Be/H is $10^{-18}$ -- $10^{-17}$
(Thomas \etal 1993, Delbourgo-Salvador \& Vangioni-Flam 1994).
  Also of interest, is the ratio $\log$(Be/Fe)
vs. [Fe/H] as is shown in Figure 2. Adopting a solar value of 
$\log$(Fe/H) = -4.465, the weighted mean of the  values in
Figure 2 (again for [Fe/H] $<$ -1) is $\log$(Be/Fe) = -5.84 $\pm$
0.05.

The boron data is taken from Duncan \etal (1997) and Garcia-Lopez \etal
(1998)  and is also shown in Figure 1.  For those stars in which Be
observations can be found, stellar parameters were again chosen
uniformly. A fit to the (NLTE) boron data for [Fe/H] $<$ -1 yields
\beq
\log (B/H) = (-9.50 \pm 0.17) + (0.67 \pm 0.09) [Fe/H]
\label{b}
\eeq
This fit is actually somewhat flatter than what one would expect
due to a simple primary explanation of the origin of B and is due
to the two somewhat discrepant points at the lower metallicities. These
points show a higher B abundance in part due to the NLTE corrections at
low metallicity (Kiselman 1994, Kiselman \& Carlsson 1996).
 Figure 3  shows the ratio B/Be as a function of [Fe/H]
taking Be abundances from the previously described compilation.  
Because of the low statistics and because of the relatively large errors
in the ratio B/Be determining a mean value from the data is difficult.
Converting an average of the log values of B/Be gives
$\langle$B/Be$\rangle
\simeq 20 \pm 4$, whereas a straight average of the unlogged ratio gives
$\langle$B/Be$\rangle \simeq 16 \pm 3$.  Alternatively, if one assumes 
that the departure from a linear relationship (in their logs)
between B,Be and Fe is simply statistical, assuming a linear fit to the data,
gives B/Be $\simeq 26 \pm 21$.  The data at present is clearly open to 
interpretation and more data particularly boron data is needed.
As we will argue below more data of both B and Be at low metallicity 
([Fe/H] $<$ -3) is needed to learn more about the origin of elements. Thus,
in the following we will consider B/Be ratios of 10, 20, and 30.

\section{Models and Results}

As described above, we compare the behavior of Be and B in the two
scenarios selected. To this aim, we exploit differences in both the
production processes and astrophysical sites. To summarize, the main
differences are as follows.  The mass domain for stars participating in
BeB nucleosynthesis are 60 -- 100\msun~ for model (a). This mass range is 
related to the acceleration of gaseous elements in superbubbles.  For
models (b), the mass range is  8 -- 100\msun~ in relation to
the acceleration of grain debris by individual SNII. The variations in 
composition and spectra are parameterized by the three values of the B/Be
ratio. This last parameter (especially for B/Be = 30)  also takes
into account the possibility of \b11 production by neutrino spallation. 

 In the framework of galactic evolution, model (a) combines two
 components, the primary LEN playing a major role in the early Galaxy,
plus the secondary standard GCR being influential in the galactic disk,
with a composition reflecting that of the ISM. The intensity of both
mechanisms at  each time step is taken to be proportional to the SN rate. 
Model (b) has a continuous primary component which has a constant
source composition. The determining difference with respect to the
standard GCR is its  primary nature since the composition of the
fragmenting nuclei emanates directly from SN and not from the ISM.

In this study we use the formalism developed in Vangioni-Flam \etal
(1996). In that work, three components playing a role in the production
of the LiBeB elements were considered.  To obtain the linear relation 
between $\log$(Be/H) and [Fe/H] (similarly for B/H), a LEN source
from stars more massive than 60 \msun~ was included (C and O interacting
on ambient H and He). Standard galactic cosmic ray (GCR)
nucleosynthesis  (fast p's and $\alpha$'s on ambient
CNO) was added. Finally, $\nu$-process nucleosynthesis
was taken into account,  albeit at a rate less than predicted in the models of
Woosley \& Weaver (1995) to adjust the \b11 to \b10 ratio and avoid \li7 
 overproduction in the early Galaxy. 
Models (a) and (b) are followed in the same way, except that the standard
GCR is absent in model (b) and that its mass domain is extended. Here, the
neutrino component is included simply through the varying B/Be ratio.
   
Before we arrive at our results, a few remarks are useful concerning the
neutrino spallation process described in Vangioni-Flam \etal (1996). In
that work, a pronounced effect on both the ratios of \b11/\b10 and B/Be
due to $\nu$-spallation was discussed.
 In both of these ratios, a bump at [Fe/H] $\approx$
-2, was predicted.  The bump was due to the fact that $\nu$-process
production of \b11 which occurs in all stars more massive than about 8
\msun~ is superimposed on other components. At very low metallicity, the
contribution of the LEN component is dominant, so it imposes its own
B/Be and \b11/\b10 ratios. Subsequently, the bulk of neutrino spallation
happens later due to the time delay correlated to the lifetime of the
stars with M $< $ 60\msun. The result is an increase of both ratios, due
to the fact that Be and \b10 are not produced in the neutrino process
while \b11 is copiously produced. If the nucleosynthesis of these
isotopes was limited to  these  two processes, there would be a fixed
value for the isotopic ratios at [Fe/H] $\gsim -2$. But, as the star
formation rate is  continuously declining, GCR nucleosynthesis, which is
included as a separate component, becomes predominant when CNO in the ISM
reaches significant abundances. Since its proper production ratios are
lower than the other components, B/Be and
\b11/\b10 decreases, and a bump appears.

 We note that the $\nu$-process 
was also predicted to have an effect on the B/Be ratio for standard 
GCR nucleosynthesis due the combination of a primary source for \b11 and
only secondary sources for \b10 and Be (Olive \etal 1994, Fields,
Olive \& Schramm, 1995). While there is no data
for \b11/\b10 at  low metallicities, it was argued by Duncan \etal (1997)
that the lack of evidence for a bump in the B/Be data (shown here in
Figure 3) minimizes a  $\nu$-process contribution to \b11. 
In model (b) however, since the mass domain involved in the production of 
all isotopes of interest is 8 -- 100\msun, it is clear that the B/Be and
\b11/\b10 ratios are constant over the galactic lifetime whether or not
neutrino spallation is operating. This model freezes out all of the light
isotopic and elemental ratios from start, so it is strongly constrained
specifically by the solar abundances. Note that neutrino spallation is a
necessity in model (b) to get the correct \b11/\b10  at the solar epoch.
In this model, a bump in B/Be or\b11/\b10 is not predicted.

One should bear in mind however, the
data are scarce and somewhat dispersed. A neutrino contribution of the
order of 30 \% cannot be excluded even in model (a). A
constraint concerning neutrino spallation could come from \li7 which
should  not be overproduced at low metallicity in order to save the
Spite plateau  (see Figure 1 in Vangioni-Flam \etal 1996). 
Model (a), thanks to its combined production processes does not 
require neutrino spallation to fit the data, although in order to 
account for the meteoritic B isotopic ratio the LEN component must 
be confined to very low particle energies, requiring that large amount 
of energy be supplied to the LEN (Ramaty et al. 1997). 
Given a dramatic improvement in the data for
B/Be, a constant ratio as a function of [Fe/H], would imply either model
(a) with a greatly reduced $\nu$-spallation contribution, or a model such
as (b).

Returning to our main goal, we now try to ascertain to what
extent the models can be tested by the existing and future data.
To this end, we will consider models in which the site for the production 
of fast C and O nuclei are SNII, in the mass range 8 -- 100 \msun~ 
(model b) and in the mass range 60 -- 100 \msun~ (model a) since the mass 
 domain is in fact the most discriminating characteristic.
We will also consider variations in the iron yield.
Generally we have assumed that the iron yield is 0.07 \msun~ over the
entire range 8 -- 100 \msun~  as observed in SN 1987 A.  Alternatively,
we will assume a yield 
of 0.07 \msun~ of Fe below 20 \msun, and  a yield proportional to the 
progenitor mass (Woosley \& Weaver 1995) at higher masses.  It is clear
that at [Fe/H] $< -1$ the Fe contribution of SNIa is insignificant. We
adjust also the B/Be ratio (at a fixed value of Be) to 10, 20, 30
corresponding to a range of possible compositions and energy spectra of
the primary component in agreement with the observations (\S 2).

One can also vary the parameters of galactic chemical evolution
such as the initial mass function (IMF) or the star formation rate (SFR).
We have chosen a simple power law form for the IMF $\phi(m) \propto
m^{-2.7}$ which is appropriate for massive stars  (Scalo 1986, Kroupa \&
Tout 1997). Lowering the slope to a Salpeter value of 2.35, changes the
overall number of massive stars and perhaps minimizes the GCR
contribution to Be production 
in model (a).  We will not consider a variation in the
IMF any further here. We consider  a SFR which is proportional to the 
gas mass fraction, $\psi = 0.3 \sigma$.  Varying the SFR would affect the
evolution of B, Be, and Fe with respect to time,  however, here we are
only considering the evolution of B and Be with respect to Fe and the
exact form of the SFR is unimportant. Our reference model is taken from 
Vangioni-Flam \etal (1996) and assumes a constant iron yield, 
LEN from only the most massive stars ($>$60 \msun~)
with B/Be set at 30. Departures from the assumption of a closed box
evolutionary model will be considered elsewhere (Fields \etal 1998).

The evolution 
of Be/H and B/H are displayed in Figures 1 and 4 and the ratio Be/Fe in
Figure 2. The solid curves in all of the figures correspond to our
reference in model (a) described above.  While it is clear that this
model  adequately describes the Be/H data, it falls short of 
the two lowest metallicity B/H observations which have been
corrected due to NLTE effects (Kiselman 1994, Kiselman \& Carlsson 1996,
 Garcia-Lopez \etal 1998).  
The corrections in these two points are about 0.8 and 0.9 dex, i.e. nearly
an order of magnitude in abundance.

The effect of decreasing the lower limit to the
progenitor mass down
to 8  \msun, is shown in Figures 1 and 2 by the dotted curves.
The net consequence is a reduced Be and B abundance at early times
(low [Fe/H]).  In terms of the ratio Be/Fe, this corresponds to 
a flat evolution vs [Fe/H] since now Be and Fe are produced in the 
same stars. This can be explained as follows:
In both cases, iron is produced by all SNIIs (8 -- 100\msun). If one
assumes that Be is only produced by the most massive stars 
(model a), the cumulative
production of Be is concentrated in a narrower range of masses, and then,
each relevant SNII should produce more BeB per Fe nucleus ejected
than in the other case (where the Be production is from 8 -- 100
\msun). Since these stars are short lived, they give rise to a higher
Be/Fe, at the very beginning of the galactic evolution. 
The required increase in 
the BeB yield per ejected Fe nucleus has energetic consequences that 
have been explored by Ramaty et al. (1997).
   
Hopefully, the predicted different evolutionary curves
will be tested in the near future but at
present, given the present range of data, [Fe/H] $>$ -3, it is not
possible to distinguish between these different sets of assumptions. More
observations at lower metallicity ([Fe/H] $<$ -3) are needed. Note that
the  effect of shortening the mass  range for the sources of the LEN
flattens the evolution curves at low metallicity 
(solid curve in 
Figure 1).  This begins to have
the appearance of a plateau-like evolution.  Therefore, if a plateau in
Be  and B vs Fe shows up observationally, one could not necessarily
conclude that 
Big Bang
nucleosynthesis (Orito \etal 1997 and references therein) took place
under inhomogeneous conditions in the early Universe.

Also shown in Figures 1 and 2 are the effects of varying the 
iron in massive stars as described above (Woosley \& Weaver 1995).
Shown in these figures by the dashed lines are the results of varying
the iron yield in model b) and thus should be directly compared with
the dotted curves. Due to the the normalization at [Fe/H] = 0, 
we find a shift in the curves relative to the case of a constant iron
yield in  model b).
 This is because massive stars give off more Fe  for a fixed amount of
Be and B produced. As expected this corresponds to a diminished 
Be/Fe ratio as seen in Figure 2. Also shown in Figure 2 is the effect of
the varying iron yield in model a).  The effect is the same, and more
importantly still allows allows for the distinction between models a)
and b) at very low metallicity. The dispersion in the data and the lack of
very low metallicity data  make it difficult at this time to distinguish
between the various models. 

Finally, we have considered the effects of the variation of the B/Be
ratio in the LEN models.  In the reference model, the B/Be ratio was
chosen to be  30.  In Figure 4, we show the results for $\log$(B/H) vs.
[Fe/H] for B/Be = 30 (solid curve), 20 (dotted curve) and 10 (dashed
curve). Given the uncertainties in the data, all of these choices must be
deemed  consistent (with the same caveat concerning the two lowest
metallicity  points described above). The case with B/Be = 10 can of
course more easily accommodate a higher rate of GCR nucleosynthesis but
this would lead to a modification to the late evolution of Be or an
increased contribution from $\nu$-spallation (cf. Vangioni-Flam \etal
1996).
\newpage
\section{Conclusion}

Be and B nucleosynthesis by 
galactic cosmic rays accelerated out of the interstellar medium
predicts a secondary
origin for Be and B so that they evolve with the square of the
metallicity. In contrast, the data have shown that these elements track
the iron  abundance linearly and therefore require a new source for their
production. Two possible primary components have been compared, a low
energy nuclear component accelerated 
in galactic superbubbles (M $> 60$
\msun, model a) 
and  C and O from grain debris accelerated by supernova
shock waves (M $>8$ \msun,  model b).
We predict a different behavior for Be/H and B/H at [Fe/H] $< -3 $
according to the domain of masses of the Be producers. For stars between
8 -- 100 \msun~ there is a strict linearity whereas, if the mass range is
reduced (60 -- 100 \msun) there is a flattening of the evolution curve
at  very low metallicity. The existing Be and B data  are presently not
sufficient to distinguish between  the various models. New data at low
metallicities, [Fe/H] $<$ -3, are essential for this purpose. 
Finally, the restricted mass range of the LiBeB stellar progenitor
(case a)  flattens the evolutionary curve and it  appears 
plateau-like.  This trend could interfere with a possible
interpretation in
 terms of inhomogeneous big bang nucleosynthesis.

This work was supported in part by DOE grant
DE--FG02--94ER--40823 at the University of Minnesota, and by PICS
no. 319, CNRS at the Institut d'Astrophysique de Paris.

\beginapjbib


\bibitem Bloemen H. \etal. 1994, A\&A, 281, L5

\bibitem Bloemen H. \etal. 1997, ApJ, 475, L25

\bibitem Boesgaard A. \& King J.R. 1993, AJ, 106, 2309

\bibitem Bykov, A.M. 1995, Space Sci. Rev., 74, 397

\bibitem Cass\'{e}, M., Lehoucq, R. \& Vangioni-Flam, E. 1995, Nature,
373, 318

\bibitem Chaussidon, M. \& Robert, F. 1995, Nature, 374, 337

\bibitem Delbourgo-Salvador, P. \& Vangioni-Flam, E. 1994, in "Origin and
 Evolution of the Elements" Edts Prantzos, N., Vangioni-Flam, E. \& Cass\'e
 , M. Cambridge University Press, p. 132

\bibitem Duncan D., Lambert D. \& Lemke M. 1992, ApJ, 401, 584

\bibitem Duncan, D., Primas, F., Rebull, L.M., Boesgaard, A.M.,
Delyiannis, C.P., Hobbs, L.M., King., \& Ryan, S.G. 1997, ApJ 488, 338

\bibitem Edvardsson, B., \etal 1994, A\&A, 290, 176

\bibitem Fields, B.D., Olive, K.A., \& Schramm, D.N. 1995, ApJ, 439, 854

\bibitem Fields, B.D., Olive, K.A., Vangioni-Flam, E., \& Cass\'e, M.
1998, in preparation

\bibitem Feltzing, S., \& Gustaffson, B. 1994, ApJ, 423, 68

\bibitem Fuhrmann, K., Axer, M., \& Gehren, T. 1993, A\&A, 271, 451

\bibitem Garcia-Lopez, R.J., Severino, G., \& Gomez, M.T. 1995,
 A\&A, 297, 787

\bibitem Garcia-Lopez, R.J. \etal 1998, ApJ, in press, Astrophys-ph/9801167 

\bibitem Gilmore, G., Gustafsson, B., Edvardsson, B. \& Nissen, P.E. 1992,
Nature 357, 379


\bibitem Kiselman, D. 1994, A\&A, 286, 169

\bibitem Kiselman, D. \& Carlsson, M. 1996, A\&A, 311, 680

\bibitem Kroupa, P. \& Tout, C.A. 1997, M.N.R.A.S., 287, 402

\bibitem Lemoine, M., Vangioni-Flam, E. \& Cass\'e, M. 1998, ApJ in press

\bibitem Lingenfelter, R. E., Ramaty, R., \& Kozlovsky, B. 1998, ApJ 
(Letters), 500, in press

\bibitem Meneguzzi, M., Audouze, J., \& Reeves, H. 1971, A\&A, 15, 337 

\bibitem Meneguzzi, M., \& Reeves, H. 1975, A\&A, 40, 99

\bibitem Meynet, G. \etal 1994, A\&AS, 103, 97

\bibitem Molaro, P., Bonifacio, P., Castelli, F., \& Pasquini, L. 1997,
 A\&A, 319, 593 

\bibitem Olive, K.A., Prantzos, N., Scully, S., \& Vangioni-Flam, E. 1994,
ApJ, 424, 666

\bibitem Orito, M., Kajino, T., Boyd, R.N. \& Mathews, G.J 1997, ApJ, 488, 515

\bibitem Pagel, B. 1991, Nature, vol 354, p. 267

\bibitem Parizot, E.M.G., Cass\'e, M. \& Vangioni-Flam, E. 1997, A\&A, 328, 107 

\bibitem Primas, F. 1995, Ph.D. thesis

\bibitem Ramaty, R., Kozlovsky, B., \& Lingenfelter, R. E. 1996, ApJ,
456, 525

\bibitem Ramaty, R., Kozlovsky, B., Lingenfelter, R.E.  \& Reeves, H.
1997, ApJ, 488, 730

\bibitem Ramaty, R., Kozlovsky, b., \& Lingenfelter, R. E. 1998,
Physics  Today, 51, no. 4, 30

\bibitem Rebolo, R., Molaro, P., Beckman J.E. 1988, A\&A, 192, 192

\bibitem Rebolo \etal 1993, in {\it Origin
 and Evolution of the Elements} eds. N. Prantzos,
 E. Vangioni-Flam \& M. Cass\'{e}
 (Cambridge University Press), p. 149

\bibitem Reeves, H. 1994, Rev.Mod.Phys., 66, 193

\bibitem Ryan, S., Bessell, M., Sutherland, R.\& Norris, J. 1990, ApJ, 348, L57
 
\bibitem Ryan, S., Norris, I., Bessel, M. \& Deliyannis, C. 1994, ApJ,
388, 184

\bibitem Samland, M. 1998, ApJ, 496,155

\bibitem Scalo, J. 1986, Fund. Cosm. Phys., 11, 1 

\bibitem Thomas, D., Schramm, D.N., Olive, K.A., \& Fields, B. 1993,
ApJ, 406, 569

\bibitem Thorburn, J.A. \& Hobbs, L.M. 1996, AJ, 111, 2106

\bibitem Vangioni-Flam, E., Cass\'e, M., Audouze, J., \& Oberto, Y. 1990, ApJ,
 364, 568  

\bibitem Vangioni-Flam, E., Lehoucq, R. \& Cass\'e, M. 1994, in "The Light
 Elements Abundances", Edts P. Crane, Springer, p.  389

\bibitem Vangioni-Flam, E. \& Cass\'e, M. 1996, in "Cosmic Abundances", Edts
 S.S. Holt \& G. Sonneborn, A.S.P. Conference series, p. 366

\bibitem Vangioni-Flam, E., Cass\'{e}, M., Fields, B.D., \& Olive,
K.A. 1996, ApJ, 468, 199

\bibitem Vangioni-Flam, E., Cass\'e, M. \& Ramaty, R. 1997, in "The Transparent
 Universe", 2nd INTEGRAL Workshop, Edts C. Winkler, T.J.L. Courvoisier \&
 Ph. Durouchoux , ESA Sp-382, p. 123

\bibitem Walker, T.P., Steigman, G., Schramm, D.N., Olive, K.A., \&
Fields, B. 1993, ApJ, 413, 562

\bibitem Woosley, S.E., Hartmann, D., Hoffman, R., Haxton, W. 1990, ApJ.,
356, 272

\bibitem Woosley, S.E. \& Weaver, T.A. 1995, ApJS,  101, 181

\endapjbib

\section*{Figure Captions}

 {\bf Figure 1:} The evolution of $\log$(Be/H) and  $\log$(B/H) with
respect to [Fe/H]. The data points are for beryllium and boron have been
described in the text. The solid curves represent the reference model
discussed in \S 3, corresponding to a mass range of Be producers between
60 -- 100 \msun, model a). The dotted curve shows the resulting evolution
when the source of BeB is extended down to 8 \msun, model b).  The dashed
curve shows the effect of varying the iron yields with the stellar mass
(Woosley \& Weaver 1995) in model b).

{\bf Figure 2:} As in Figure 1 for the evolution of $\log$(Be/Fe) as a
function of [Fe/H]. Also shown by the dot dashed curve is the case of a
variable iron yield in model a).

{\bf Figure 3:} The data for B/Be as a function of [Fe/H].

 {\bf Figure 4:} As in Figure 1, the evolution of $\log$(Be/H) and
$\log$(B/H) with respect to [Fe/H].
The solid curves represent the reference model discussed in the text.
Also shown is the resulting evolution when the B/Be yield is fixed
at 20 (dotted curve) and at 10 (dashed curve).


\newpage

\begin{figure}[htb]
\vskip 1in
\epsfysize=7truein
\epsfbox{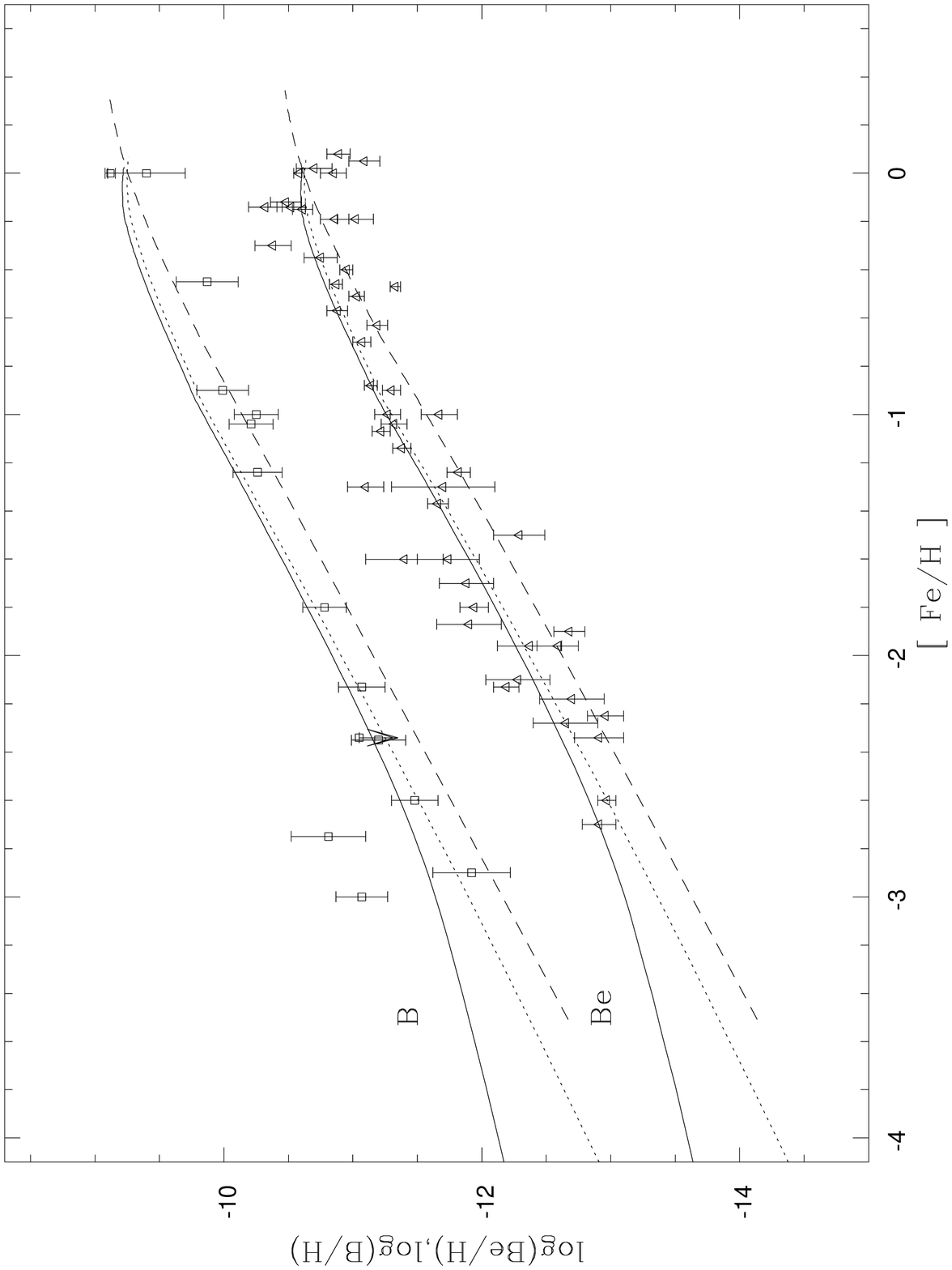}
\end{figure}  

\newpage

\begin{figure}[htb]
\vskip 1in
\epsfysize=7truein
\epsfbox{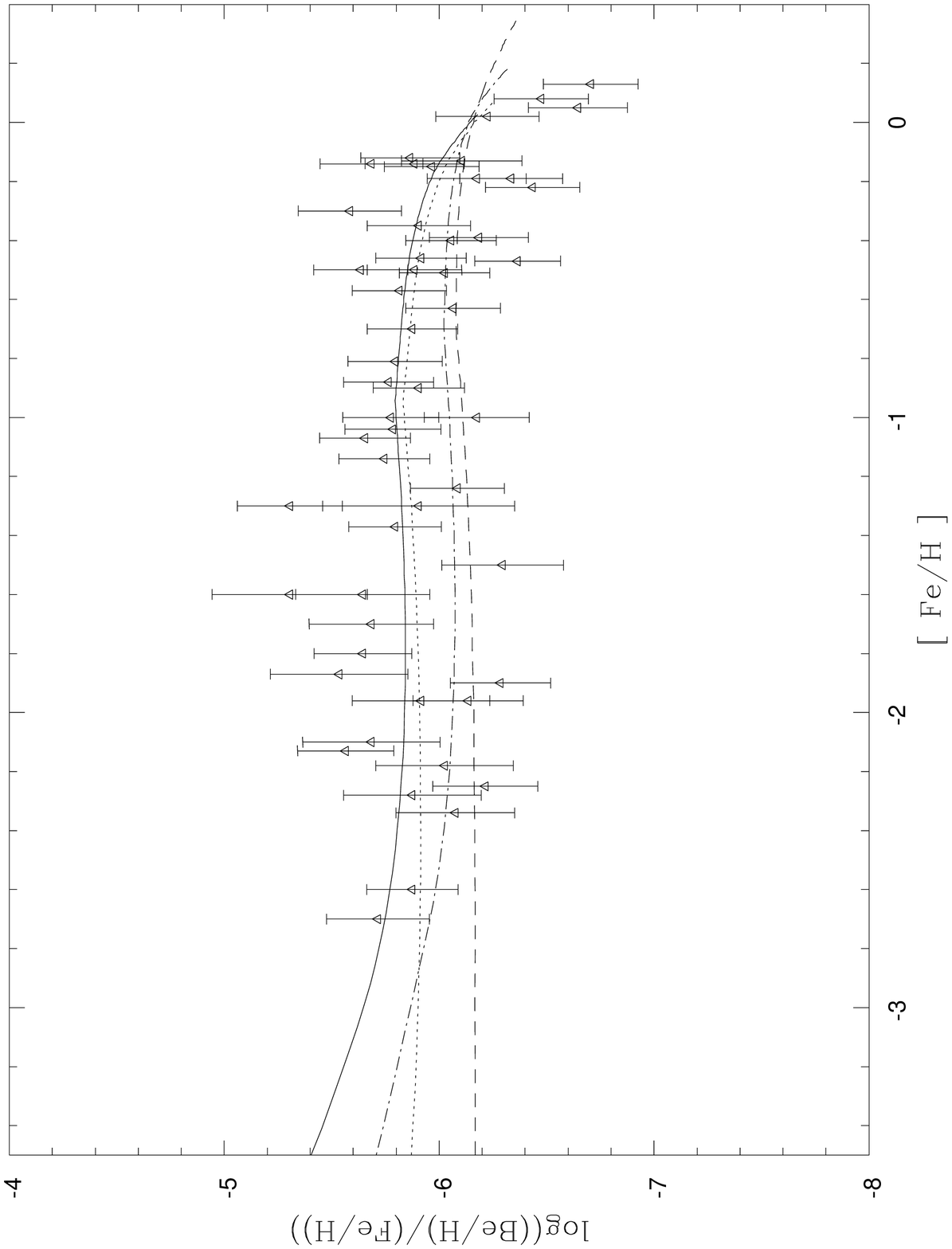}
\end{figure}

\newpage

\begin{figure}[htb]
\vskip 1in
\epsfysize=7truein
\epsfbox{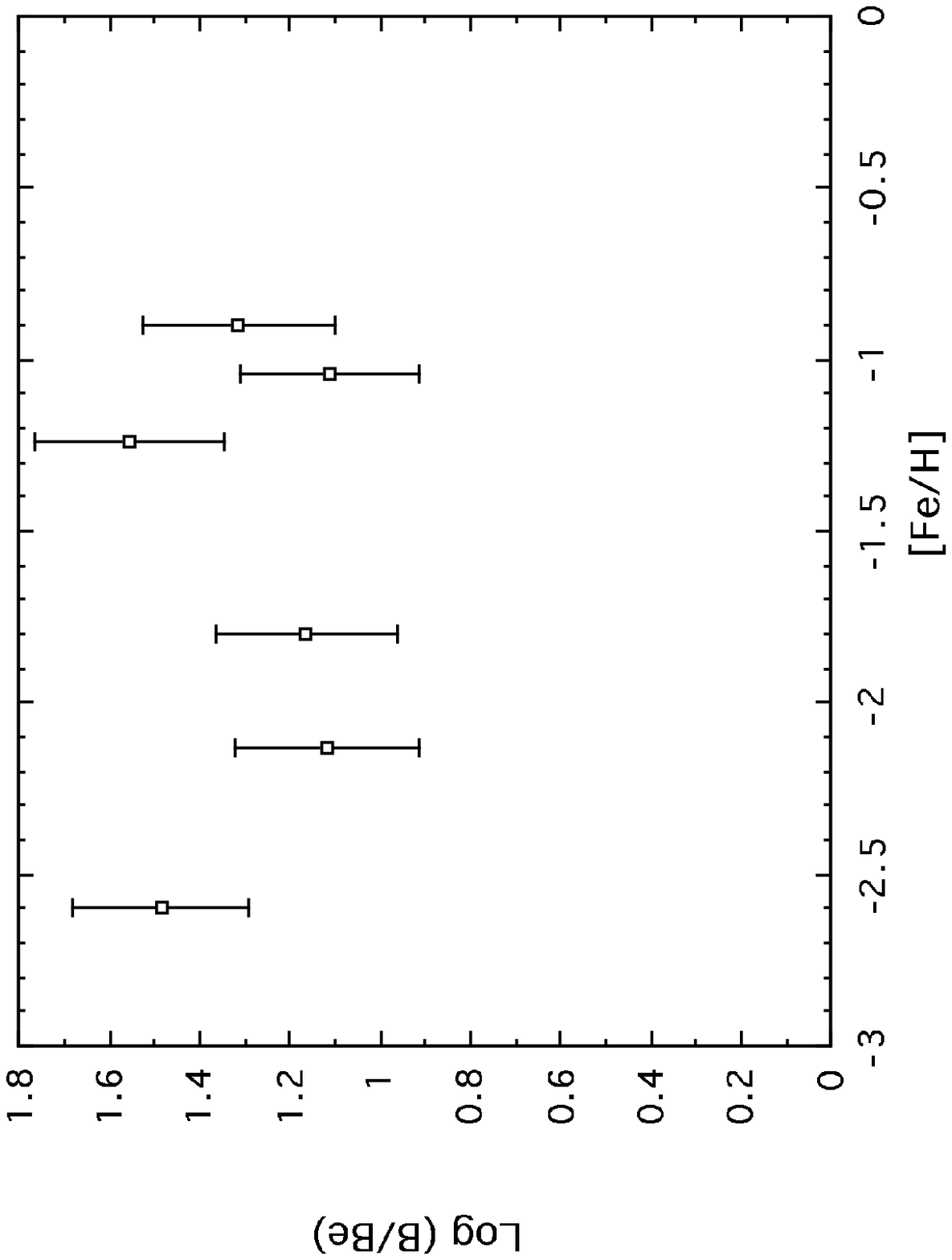}
\end{figure}  

\newpage

\begin{figure}[htb]
\vskip 1in
\epsfysize=7truein
\epsfbox{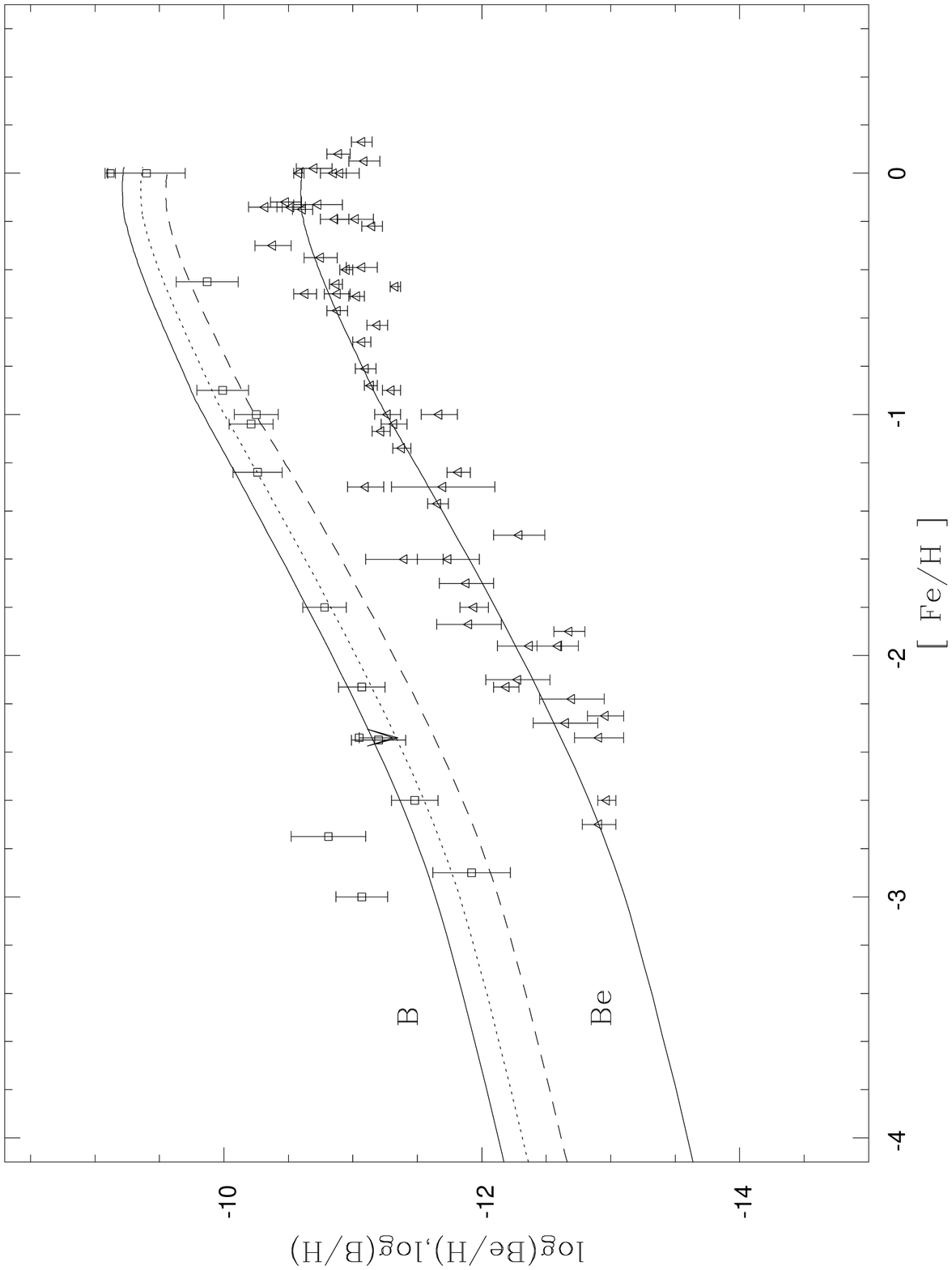}
\end{figure}

\end{document}